\input epsf
\documentstyle[aps]{revtex}
\begin{document}
\title{Lamb shift calculated by simple noncovariant method}

\author{Guang-jiong Ni\thanks{E-mail: Gjni@fudan.ac.cn}}
\address{Department of Physics, Fudan University, Shanghai, 200433,
  P. R. China}
\author{Jun Yan\thanks{E-mail: jy272@scires.nyu.edu}}

\address{Department of Physics, New York University, 4 Washington
  Place, New York, NY, 10003}

\maketitle

\begin{abstract}
The Lamb Shift (LS) of Hydrogenlike atom is evaluated by a simple method of
quantum electrodynamics in noncovariant form, based on the relativistic
stationary Schr\"odinger equation. An induced term
proportional to $\overrightarrow{p}^4$ in the effective Hamiltonian is
emphasized. Perturbative calculation of second order leads to the LS of $
1S_{1/2}$ state and that of $2S_{1/2}-2P_{1/2}$ states in H atom with the
high accuracy within 0.1\% 
\end{abstract}
\section{Introduction}
\label{sec:intro}
The experimental discovery of Lamb Shift (LS) in 1947 and its theoretical
explanations that followed are of great importance for the establishment of
Quantum Electrodynamics (QED). (Ref.\cite{Sakurai}-\cite{Weinberg}). The
experimental value of the energy difference between $2S_{1/2}$ and $2P_{1/2}$
states for Hydrogen atom reads (in unit of microwave frequency)

\begin{equation}
  \label{H2S2P}
  E(2S_{1/2})-E(2P_{1/2})=1057.845MHz
\end{equation}
while the absolute LS for $1S_{1/2}$ state is \cite{Weitz}

\begin{equation}
\Delta E_H(1S_{1/2})=8172.86MHz  \label{H1S}
\end{equation}

\begin{equation}
\Delta E_D(1S_{1/2})=8184.00MHz  \label{D1S}
\end{equation}

The LS only accounts for the order of $10^{-6}$ or $O(\alpha ^3)$ of that of
the binding energy for electron, i.e., that of Rydberg energy

\begin{equation}
R_y=R_H=\frac {1}{2}\alpha ^2\mu =3.28805128\times 10^9MHz  \label{RH}
\end{equation}
where $\mu$ is the reduced mass of electron and $\alpha =\frac{e^2}{4\pi }
=1/137.0359895$. ($\hbar=c=1$).

The theoretical investigation over 50 years reveals
that:

(a) The main contribution of LS comes from the difference of radiative
correction (i.e., the perturbative energy stemming from emitting 
a virtual photon and then absorbing it) in different states of electron.

(b) The difference between the wave functions of $S$ states and $P$ states is
important. The electron in $S$ states has more probability to move into the
vicinity of nucleus. In other words, it has more high momentum components in
the momentum representation of wave function for $S$ states.

In some literature, in the integration of momentum $k$ of virtual photon,
the range of $k$ was often divided into two regions. For low $k$ from $k=0$
to, say, $k=\alpha m_e=\alpha m$, the binding effect of electron is taken into account in
noncovariant theory, whereas for high $k$ up from $k=\alpha m$ the covariant theory
of QED is applied. This kind of treatment seems to us is difficult to avoid
the double counting in virtual electron states conceptually. Moreover, the
so-called long wave (i.e., $E1$) approximation was used in the noncovariant
theory at low $k$ region as $e^{ikr}\sim 1$. But $exp(ikr)\sim
exp(i\alpha m a)\sim exp(i)\sim 1$ ($a=1/\alpha \mu$ being the Bohr
radius) is also doubtful to be a good approximation.

We wish to restudy the problem in noncovariant scheme. First of all, the
mystery of LS is not only related to the small scale of energy
shift shown at Eqs.(\ref{H2S2P})-(\ref{D1S}) but also to the high accuracy
of the noncovariant calculation based on the Stationary Schr\"odinger Equation (SSE):

\begin{equation}
H_0\psi =\varepsilon \psi  \label{SSE}
\end{equation}
with 
\begin{equation}
H_0=\frac{p^2}{2\mu }-\frac{Z\alpha }{r}  \label{H0}
\end{equation}
and 
\begin{equation}
\varepsilon =-\frac{Z^2\alpha ^2\mu }{2n^2}  \label{e}
\end{equation}

According to the theory of special relativity (SR), the energy of free
electron reads 
\begin{equation}
E=\sqrt{\mu ^2+p^2}=\mu +\frac{p^2}{2\mu }-\frac{p^4}{8\mu ^3}+\cdots  \label{E}
\end{equation}
The magnitude ratio of the third term to the second one is approximately $
\langle \frac {p^2}{4\mu ^2} \rangle \sim \langle \frac {p^2} {2\mu} \rangle
\frac {1} {\mu} \sim \frac {\left| \varepsilon \right|} {\mu} \sim O(\alpha ^2)$ in a
Hydrogenlike atom. So at first sight, the relativistic modification to $H_0$
would be expected to account for an energy decrease of $\varepsilon $ to
order of $10^5MHz.$ However, in fact, the LS shows an energy increase of S
states to order of $10^3MHz$ only. This is a mystery we should consider
first before the LS could be understood in the noncovariant formalism of QED.

\section{Relativistic SSE}
\label{sec:RSSE}
In Refs.\cite{NIChen1997}-\cite{NIpreprint97}, it was argued that the SSE,
Eq.(\ref{SSE}), is essentially relativistic as long as the eigenvalue $
\varepsilon $ is related to the binding energy $B$ as follows:

\begin{equation}
B=Mc^2[1-(1+\frac{2\varepsilon }{Mc^2})^{1/2}]  \label{binding}
\end{equation}
where $M=m+m_N$ is the total mass of Hydrogenlike atom with $m_N$ being the
mass of nucleus. Obviously, when $\varepsilon \ll Mc^2$, $B\simeq
-\varepsilon $ as expected.

Eqs.(\ref{SSE})-(\ref{e}) together with Eq.(\ref{binding}) was derived from the
time-dependent Schr\"odinger equation of two-body (say
electron-proton) system in combination with a basic symmetry 
\begin{equation}
\theta (-\overrightarrow{r_e},-\overrightarrow{r_p},-t)=\chi (
\overrightarrow{r_e},\overrightarrow{r_p},t)  \label{transition}
\end{equation}

The explanation is as follows. The electron and proton are all not pure.
They not only have a particle state $\theta (\overrightarrow{r_e},
\overrightarrow{r_p},t)$, but also have a hiding antiparticle state $\chi (%
\overrightarrow{r_e},\overrightarrow{r_p},t)$, $\theta $ and $\chi $ are
coupled together according to the symmetry (\ref{transition}). For one body
system, this symmetry leads to Klein-Gordon equation (without spin) or Dirac
equation (with spin) (Refs.\cite{NiChen1996} \cite{NiZhouYan}). The symmetry
that ``the space-time inversion is equivalent to particle-antiparticle
transformation'' shown as (\ref{transition}) is the essence of SR. The
various strange effects of SR are nothing but the reflections of antimatter
which is in a subordinate status ($\left| \chi \right| <\left| \theta
\right| )$ and is just displaying its presence tenaciously\cite{NiKexue1}.

Therefore, at the level of quantum mechanics, the Hamiltonian $H_0$ in SSE
Eq.(\ref{SSE}) is enough in the form of Eq.(\ref{H0}). No $p^4$ term like
that in Eq.(\ref{E}) is needed for relativistic correction. The latter is
brought into consideration implicitly at the final stage $\varepsilon
\rightarrow -B$ as shown by Eq.(\ref{binding}).

\section{Semiempirical calculation of LS}
\label{sec:SCLS}
Based on SSE, we consider that in an effective Hamiltonian a term $b_2p^4$
will be induced at QED level by radiative correction. It will lead to an
energy shift of Hydrogenlike atomic level as:

\begin{equation}
\Delta E_{Z,nl}^{Rad} =\langle Znl\left| b_2p^4\right|
Znl\rangle
=[\frac{8n}{(2l+1)}-3]\frac{b_2Z^4}{n^4a^4} \hspace{0.5in} (a=\frac 1{\alpha \mu })  \label{empirical}
\end{equation}

The coefficient $b_2$ can be fixed from experiment by the LS
between $2S_{1/2}-2P_{1/2}$ states in Hydrogen:

\begin{equation}
\frac{b_2}{a_H^4}=\frac {3}{2}(1057.845-0.087)MHz=1586.637MHz  \label{B2}
\end{equation}
where $0.087MHz$ is the small correction stemming from the finite nucleus
radius.

To calculate the absolute LS of $1S$ state, Eq.(\ref{H1S}), we should add
three extra contributions:

(a) The vacuum polarization in QED induces a decrease in charge of electron: \cite{NiWangYanLi}, \cite{NiWang}
\begin{equation}
\Delta \alpha =-\frac{\alpha ^4Z^2}{3\pi n^2}  \label{alpha}
\end{equation}
which leads to an increase of energy 
\begin{equation}
\Delta E_{Z,nl}^{Vp}=\frac{2Z^4\alpha ^3}{3\pi n^4}R_y  \label{Vp}
\end{equation}
For Hydrogen 1S state, it reads 
\begin{equation}
\Delta E_{H,1S}^{Vp}=271.140MHz  \label{HVp}
\end{equation}

(b) The relativistic correction from Eq.(\ref{binding}) reads 
\begin{equation}
\Delta E^{Rel}=-\frac{\varepsilon ^2}{2M}=-\frac 14\alpha ^2\frac \mu M\frac{
R_y}{n^4}  \label{Rel}
\end{equation}
which yields a decrease in energy: 
\begin{equation}
\Delta E_{H,1S}^{Rel}=-23.814MHz  \label{HRel}
\end{equation}

(c) The correction from finite nucleus radius ($r_N$) reads

\begin{equation}
\Delta E_{nl}^{Nu}=\frac 45\frac 1{n^3}(\frac{r_N}a)^2R_y\delta _{l0}
\label{Nu}
\end{equation}
which contributes a small increase of energy 
\begin{equation}
\Delta E_{H,1S}^{Nu}=0.697MHz  \label{HNu}
\end{equation}
Altogether, we obtain theoretically 
\begin{equation}
\Delta E_{H,1S}^{Theory}=8181.208MHz  \label{Htheory}
\end{equation}
The deviation between (\ref{Htheory}) and the experimental value, Eq.(\ref
{H1S}) is only 0.1\%.

\section{Calculation of $b_2$ from the first principle}
\label{sec:1st principle}
We are now in a position to derive the value of coefficient $b_2$ from the
first principle of QED in noncovariant form. Consider an electron with
charge $-e$ is moving with (three dimensional) momentum $\overrightarrow{p}$ in
the center of mass system of Hydrogenlike atom and is coupled to the
electromagnetic field via two kinds of interactions (\cite{Sakurai}, \cite{Davydov}) 
\begin{equation}
H^{(1)}=\frac {e}{\mu c}\hat {\overrightarrow{A}}\cdot \hat {\overrightarrow{p}}  \label{H1}
\end{equation}
\begin{equation}
H^{(2)}=\frac{e\hbar }{2\mu c}\overrightarrow{\sigma }\cdot \overrightarrow{
\nabla }\times \hat{\overrightarrow{A}}  \label{H2}
\end{equation}

According to the perturbation theory in quantum mechanics, the energy shift
due to (\ref{H1}) from original $\varepsilon _p=\frac{\overrightarrow{p}^2}{2\mu }$ will be 
\begin{equation}
\Delta E_p^{(1)}=\sum_{i}
\frac{\left| \left\langle i\left|
H^{(1)}\right| \overrightarrow{p}\right\rangle \right| ^2}{\varepsilon
_p-\varepsilon _i}  \label{energyshift1 }
\end{equation}
where the stationary state $\left| \overrightarrow{p}\right\rangle =\frac
1{\sqrt{V}}e^{i\overrightarrow{p}\cdot \overrightarrow{r}}$ is normalized in
a volume $V$. The intermediate (virtual) state $\left| i\right\rangle $ is
composed of a plane wave eigenstate of SSE (Eq.(\ref{SSE}))
$\overrightarrow {q}$ and a virtual photon with continuous
momentum $\overrightarrow{k}$, see Fig. 1. So the $\varepsilon _i$ in the
denominator of (\ref{energyshift1 }) reads $\varepsilon _i=\frac{
\overrightarrow{q}^2}{2\mu }+\omega _k$, ($\omega _k=\left| \overrightarrow{k}\right| =k)$. The quantized potential $\hat{\overrightarrow{A}}$ of
electromagnetic field reads as usual: 
\begin{equation}
\hat {\overrightarrow{A}}(\overrightarrow{r},t)=\int \frac{d
\overrightarrow{k}}{(2\pi )^{3/2}}\frac 1{\sqrt{2\omega _k}}\sum_{\lambda=1,2} \overrightarrow{\epsilon }_{\overrightarrow{k},\lambda }\left( 
\hat{a}_{\overrightarrow{k}}(t)e^{i\overrightarrow{k}\cdot 
\overrightarrow{r}}+\hat{a}_{\overrightarrow{k}}^{\dagger
}(t)e^{-i\overrightarrow{k}\cdot \overrightarrow{r}}\right)   \label{A}
\end{equation}

After the integration of matrix element of $H^{(1)}$ with respect to space,
we substitute one $\delta $ function $\delta \left( \overrightarrow{p}-
\overrightarrow{q}-\overrightarrow{k}\right) $ by $\frac V{(2\pi )^3}$ and \
then perform the integration with respect to $\overrightarrow{q}$, yielding 
\begin{equation}
\Delta E_p^{(1)}=-\frac{\alpha p^2}{\pi \mu }\int_{-1}^1d\eta (1-\eta ^2)I
\label{energyshift1Integrate }
\end{equation}
\begin{equation}
I=\int_0^\infty \frac{dk}{k+\xi }  \label{Integration1}
\end{equation}
where $\eta =\frac{\overrightarrow{p}\cdot \overrightarrow{k}}{p\cdot k
}$ and $\xi =2(\mu -p\eta )$.

\section{The renormalization is a procedure to reconfirm the mass.}
\label{sec:renomalization}
As in the calculation of QED in covariant form, we also encounter the
divergent integral $I$, Eq.(\ref{Integration1}), here in the integration of
three dimensional momentum $k$ of virtual photon. To treat the divergence,
we will follow the spirit of a simple but effective method used in covariant
quantum field theory, which evolved from the so-called differential
renormalization in the literature \cite{Freedman}-\cite{Smirnov}, then was
proposed by Ji-feng Yang \cite{Yang} and applied extensively in Refs.\cite
{NiWang}, \cite{NIChen9708155}, \cite{NiLouLuYang} (see also the discussion
in \cite{NiKexue3}). Here the trick is as follows.

Take the derivative of integral with respect to the parameter $\xi $ having
a mass dimension:

\begin{equation}
\frac{\partial I}{\partial \xi }=-\int_0^\infty \frac{dk}{\left( k+\xi
\right) ^2}=-\frac 1\xi  \label{DI}
\end{equation}
which is convergent now. Then we reintegrate Eq.(\ref{DI}) with respect to $
\xi $ for returning back to $I$

\begin{equation}
I=-\ln\xi +C_1  \label{I reintegrate}
\end{equation}
where an arbitrary constant $C_1$ appears. Substituting (\ref{I reintegrate}
) into (\ref{energyshift1Integrate }), one obtains 
\begin{eqnarray}
\Delta E_p^{(1)} &=&\frac{\alpha \mu }\pi[(\frac 23(\frac p\mu )^2+\frac
p\mu -\frac \mu {3p})\ln(1+\frac p\mu )+(\frac 23(\frac p\mu )^2-\frac p\mu
+\frac \mu {3p})\ln(\left| 1-\frac p\mu \right| ) \nonumber \\
&&-\frac{16}9(\frac p\mu )^2+\frac 23+(\frac 43\ln2+\frac 43\ln\mu -\frac
43C_1)(\frac p\mu )^2]  \nonumber \\
&=&b_1^{(1)}p^2+b_2^{(1)}p^4+\cdots  \label{energyshift1 of p }
\end{eqnarray}
\begin{equation}
b_1^{(1)}=\frac \alpha {\pi \mu }\left( \frac 43\ln2+\frac 43\ln\mu -\frac
43C_1\right)   \label{b11}
\end{equation}
\begin{equation}
b_2^{(1)}=\frac \alpha {\pi \mu ^3}\left( -\frac 2{15}\right)   \label{b12}
\end{equation}

Note that, however, the term $b_1^{(1)}p^2$ will be combined into the kinetic
energy term in original $H_0$. They are indistinguishable. The appearance of
arbitrary constant $C_1$ precisely reflects the fact that we can not
calculate the reduced mass of an electron via the perturbation calculation
of $\Delta E_p^{(1)}$. So the value of $C_1$ is chosen such that $
b_1^{(1)}=0$, implying that the value of reduced mass $\mu$
(yet not $\mu_{obs} $, see below) is reconfirmed as an observed mass which can
only be fixed by the experiment (not by theory).

Next turn to $H^{(2)}$, which induces the spin flip process between $\left| 
\overrightarrow{p},\pm \frac 12\right\rangle $ and $\left| \overrightarrow{q}
,\pm \frac 12\right\rangle $ states, yielding 
\begin{eqnarray}
\Delta E_p^{(2)} &=&\frac 12\sum_{i,s_z=\pm \frac 12} \frac{
\left| \left\langle i\left| H^{(2)}\right| \overrightarrow{p}
,s_z\right\rangle \right| ^2}{\varepsilon _p-\varepsilon _i}  \nonumber \\
&=&-\frac \alpha {2\pi \mu }\int_{-1}^1d\eta J  \label{energyshift2 }
\end{eqnarray}
\begin{equation}
J=\int_0^\infty \frac{k^2dk}{k+\xi }  \label{J}
\end{equation}

For this divergent integral, derivative of third order is needed to render
it convergent: 
\begin{equation}
\frac{\partial ^3J}{\partial \xi ^3}=-\frac 2\xi  \label{DJ}
\end{equation}

So after reintegrating with respect to $\xi$, we have:

\begin{equation}
J=-\xi ^2\ln\xi +C_2\xi ^2+C_3\xi +C_4  \label{J reintegrate}
\end{equation}
\begin{eqnarray}
\Delta E_p^{(2)} &=&\frac{\alpha \mu }\pi \{\frac{2\mu }{3p}[(1+\frac p\mu
)^3\ln(1+\frac p\mu )-(1-\frac p\mu )^3\ln(\left| 1-\frac p\mu \right| )
]-\frac {22}{9}(\frac p\mu )^2-\frac {4}{3}  \nonumber \\
&&+4\left( \ln2+\ln\mu \right) -4C_2-\frac{2C_3}\mu -\frac{C_4}{\mu ^2}+\left(
\frac 43\ln2+2+\frac 43\ln\mu -\frac 43C_2\right) (\frac p\mu )^2\}  \nonumber
\\
&=&b_0^{(2)}+b_1^{(2)}p^2+b_2^{(2)}p^4+\cdots  \label{energyshift2 of p }
\end{eqnarray}
\ 
\begin{equation}
b_0^{(2)}=\frac{\alpha \mu }\pi \left[ 4\left( \ln2+\ln\mu \right) -4C_2-\frac{
2C_3}\mu -\frac{C_4}{\mu ^2}\right]   \label{b20}
\end{equation}
\begin{equation}
b_1^{(2)}=\frac \alpha {\pi \mu }\left( \frac 43\ln2+2+\frac 43\ln\mu -\frac
43C_2\right)   \label{b21}
\end{equation}

\begin{equation}
b_2^{(2)}=\frac \alpha {\pi \mu ^3}\left( -\frac 1{15}\right)  \label{b22}
\end{equation}

We shall fix three arbitrary constants $C_2$, $C_3$ and $C_4$ carefully.
First look at the $b_1^{(2)}p^2$ term which should be combined with the term 
$\frac{p^2}{2\mu}$ with $\mu$ already fixed. Any
modification on $\mu$ must be finite and fixed. So the only
possible choice of $C_2$ is to cancel the ambiguous term\ $\frac 43 \ln \mu $
in $b_1^{(2)}$, leaving 
\begin{equation}
b_1^{(2)}=\frac {\beta} {2\mu } \hspace{0.5in}\beta =\frac{2\alpha }
\pi \left( \frac 43ln2+2\right)   \label{b21fixed}
\end{equation}

The constants $C_3$ and $C_4$ must be chosen such that $b_0^{(2)}=0,$ which
means that we reconfirm the SSE as our starting point. There is always no
rest energy term in SSE.

Hence, the nonzero contribution of $b_1^{(2)}p^2$ does bring a finite and
fixed modification on $\mu$ so that 
\begin{equation}
\mu _{obs}=\frac \mu {1+\beta }
\end{equation}
where $\mu_{obs} $ is the reduced mass eventually observed in experiment. However,
as discussed in Eq.(\ref{E}), there is no term like $-\frac{p^4}{8\mu_{obs}^3}$
in SSE. The relativistic correction is left to the modification of $
\varepsilon $ to $B$ at the final stage. So an unobservable term is also
modified, the difference $-\frac 18\left( \frac 1{\mu
_{obs} ^3}-\frac 1{\mu^3}\right) p^4$ should be treated also as an unobservable
background of modification in $p^4$ term, which should be subtracted off
from the observable one. 
Hence while $b_1=b_1^{(1)}+b_1^{(2)}=b_1^{(2)}$,
but instead of $b_2=b_2^{(1)}+b_2^{(2)}=\frac \alpha {\pi \mu ^3}\left(
-\frac 15\right)$, we should have a renormalized $b_2^R$ as 
\begin{equation}
b_2^R=b_2+\frac 1{8\mu ^3}\left( 3\beta+3\beta^2+\beta^3\right) =\frac
\alpha {\pi \mu_{obs} ^3}\left( 1.942816878\right)   \label{b2r}
\end{equation}

\section{Comparison with the experimental values}
\label{sec:comparision}
Let us use Eq.(\ref{b2r}) to evaluate the LS of H atom: 
\begin{equation}
\Delta E_{H,1S}^{Rad}=\left\langle b_2^Rp^4\right\rangle _{1S}=\frac{5b_2^R}{
a_H^4}=7901.629MHz  \label{Rad 1S}
\end{equation}
Adding the contributions from Eqs.(\ref{HVp}), (\ref{HRel}) and (\ref{HNu}),
we get 
\begin{equation}
\Delta E_{H,1S}^{Theory}=8149.653MHz  \label{Theory 1S}
\end{equation}
which is smaller than the experimental value Eq.(\ref{H1S}) up to 0.28\%.

For the LS between $2S_{1/2}$ and $2P_{1/2}$ states, 
\begin{equation}
\Delta E_{H,2S_{1/2}-2P_{1/2}}^{Rad}=\frac 23\frac{b_2^R}{a_H^4}=1053.551MHz
\label{Rad 2S-2P}
\end{equation}
After adding a small contribution due to the finite nucleus radius, we have 
\begin{equation}
\Delta E_{H,2S_{1/2}-2P_{1/2}}^{Theory}=1053.638MHz  \label{Theory 2S-2P}
\end{equation}
which is also smaller than the experimental value Eq.(\ref{H2S2P}) up to
0.40\%.

For further improvement, we keep all $p^n$ ($n\ge 4$) terms. So we
manage to evaluate the renormalized radiation correction as follows 
\begin{equation}
\Delta E^{Rad}(\overrightarrow{p})=\Delta E_p^{(1)}+\Delta
E_p^{(2)}-b_1^{(2)}p^2-(\sqrt{p^2+\mu_{obs} ^2}-\mu_{obs} -\frac{p^2}{2\mu_{obs} })+(\sqrt{
p^2+\mu ^{2}}-\mu-\frac{p^2}{2\mu})
\label{RadiativeEnergyrenormalizied}
\end{equation}
with $\mu_{obs}-\mu=d\mu=-\beta\mu_{obs}$ and $\sqrt{p^2+\mu_{obs} ^2}-\sqrt{
p^2+\mu ^{2}}=\frac{\mu_{obs} d\mu }{\sqrt{p^2+\mu_{obs} ^2}}$ as we
wish to keep the explicit dependence on $\alpha$ throughout
Eq. (\ref{RadiativeEnergyrenormalizied}) being of order $O(\alpha)$. Then we calculate
the expectation value of $\Delta E^{Rad}(\overrightarrow{p})$ for a fixed
state numerically, yielding 
\begin{equation}
\Delta E_{H,1S}^{Rad}=7920.533MHz  \label{1stRad1SH}
\end{equation}
\begin{equation}
\Delta E_{H,2S_{1/2}-2P_{1/2}}^{Rad}=1057.550MHz  \label{1stRad2S2pH}
\end{equation}
\begin{equation}
\Delta E_{D,1S}^{Rad}=7922.688MHz  \label{1stRad1SD}
\end{equation}
\begin{equation}
\Delta E_{D,2S_{1/2}-2P_{1/2}}^{Rad}=1057.838MHz  \label{1stRad2S2pD}
\end{equation}
After adding the other three corrections Eqs.(\ref{Vp})-(\ref{Nu}), we are
pleased to see the results 
\begin{equation}
\Delta E_{H,1S}^{Theory}=8168.557MHz  \label{1sttheory1SH}
\end{equation}
\begin{equation}
\Delta E_{H,2S_{1/2}-2P_{1/2}}^{Theory}=1057.637MHz  \label{1sttheory2S2pH}
\end{equation}
\begin{equation}
\Delta E_{D,1S}^{Theory}=8186.181MHz  \label{1sttheory1SofD}
\end{equation}
\begin{equation}
\Delta E_{D,2S_{1/2}-2P_{1/2}}^{Theory}=1058.363MHz  \label{1sttheory2S2pD}
\end{equation}
coinciding with the experimental data to a high accuracy ($\lesssim 0.1\%$).

\section{Summary and discussion}
\label{sec:Summary}
(a) We propose a simple but effective method for calculating the LS in a
noncovariant form. Two crucial observations are as follows:

(i) The SSE is essentially relativistic as long as its eigenvalue $
\varepsilon $ is related to the binding energy $B$ by Eq.(\ref{binding}).

(ii)The main contribution to LS is coming from the different mean values of
operator $\overrightarrow{p}^4$ in $S$ and $P$ states as shown in the
semiempirical calculation. The analysis convinced ourselves that the use of
three dimensional momentum $\overrightarrow{p}$ is much suitable than that
of four dimensional one. Actually, a simple one-loop calculation in
covariant form of QED gave us the value of LS for H atom with accuracy only
5\% for $2S_{1/2}-2P_{1/2}$ states and even worse ($\sim 22\%)$ for $
1S_{1/2} $ state \cite{NiWang}.

(b) The subtlety of SSE can be seen further by the following
comparison. At free moving condition, the relativistic effect is
contained in the definition $\varepsilon=\frac {p^2}{2\mu}$ with rest
(reduced) mass being a constant containing all the radiative
correction. On the other hand, when the electron is bound in a
Hydrogenlike atom, its Hamilton $H_0$ contains no rest mass and no explicit (negative) $\overrightarrow{p}^4$ term
either. The relativistic effect is contained in the definition $\varepsilon =
\frac{E^2-M^2}{2M}$ (with $E$ and $M$ being the total energy and mass of the
system) and Eq.(\ref{binding}) implicitly. However, the radiative correction
does induce a small (positive) term of $\overrightarrow{p}^4$ at the level
of QED.

(c) The definition of $\varepsilon $ and Eq.(\ref{binding}) also indicate
that there is no any negative energy eigenstate in SSE. The hiding
antiparticle state (which is the essence of special relativity) is already
taken into account in deriving SSE with Eq.(\ref{binding}). In other words,
no further explicit virtual positron state should be considered in our
calculation. Actually, we had struggled for years before eventually
realizing that only the simple formulas (\ref{energyshift1 }) and (\ref{energyshift2 }) with Fig.1 are needed in the perturbative calculation of
second order.

(d) The effective Hamiltonian of Hydrogenlike atom for evaluating the LS can
be summarized as 
\begin{equation}
H_{eff}=\frac{p^2}{2\mu }-\frac{Z\alpha }r+\Delta E^{Rad}(\overrightarrow{p}
)+\frac{\alpha ^4Z^3}{3\pi n^2r}  \label{effHamilton}
\end{equation}
while the third term $(\simeq b_2^Rp^4)$ is the QED modification to the
first term, the fourth term could be seen as that to the second term (see
Eqs.(\ref{alpha}), (\ref{Vp})).

Eq.(\ref{effHamilton}) is applicable to $j=1$ states. The $P_{3/2}$
state is pushed up by extra spin-orbit coupling to form the fine
structure. For Hydrogen, $E(2P_{3/2})-E(2P_{1/2})=1.09691\times
{10}^{4}MHz$ about $10$ times of LS. Furthermore, the hyperfine structure
(hfs) in Hydrogen, stemming from the interaction between the magnetic
moments of electron and proton, is smaller than the LS. The energy
splitting of $1S_{1/2}$ states is well known as
$\Delta E_{1S_{1/2}}(hfs)=1420.406MHz$. As we don't take the magnetic
moment of nuclei into account, the hfs is not considered in this paper.

In deriving the $H_{eff}$, the Pauli interaction term $H^{(2)}$
between electron spin and the external magnetic field has to be added
to  the RSSE so that the spin-orbital coupling (hyperfine structure)
and LS can be calculated quantitatively. This is the price we must pay
for the use of RSSE. On the other hand, though the Dirac equation in
external field can predict the electron spin with $g=2$, it fails to
take the difference of reduced mass of electron into account because
it is a one-body equation. Furthermore, it predicts a too low ground state ($1S$)
and a too large splitting of $2P_{3/2}$ and $2P_{1/2}$ states. The
weakness of Dirac equation seems to us is due to its overestimation of
the antiparticle ingredient in the electron as discussed in
Ref.\cite{NiZhouYan}. Actually, only after long time study on Dirac
equation with its difficulty in calculating LS, especially the absolute 
LS of $1S$ states, could we believe in the advantage of using RSSE as
the starting point.  

(e) The reasonable result in this paper again shows the correctness and
effectiveness of the renormalization method used in Refs.\cite{Yang}-\cite{NiKexue3}. For treating the divergence , which warns us of the lack of our
knowledge about the parameters (mass, charge, etc.), the renormalization is
a procedure to reconfirm the parameters step by step rigorously. Here it is
interesting to see an example of finite and fixed renormalization. Note that, however, we must consider the contribution of $H^{(1)}$ first
to define the parameter $\mu$ before to consider that of $H^{(2)}$ for
bringing $\mu$ to $\mu_{obs}$. This is the only reasonable logic. The inverse
logic, i.e., to consider $H^{(2)}$ first and then $H^{(1)}$ next, would lead to
inconsistent and wrong result. 
\section*{Acknowledgments}
\label{sec:Acknowledgments}
We thank Prof. D. Zwanziger, Mr. Hailong Li and Mr. Haibin Wang for
discussions. We also thank Mr. Sangtian Liu in NYU who gave us a lot
of help in \LaTeX and figure of this paper. This work was supported
in part by the NSF of China.   
\begin{figure}[htbp]
  \begin{center}
\centerline{\epsfxsize=5cm\epsfbox{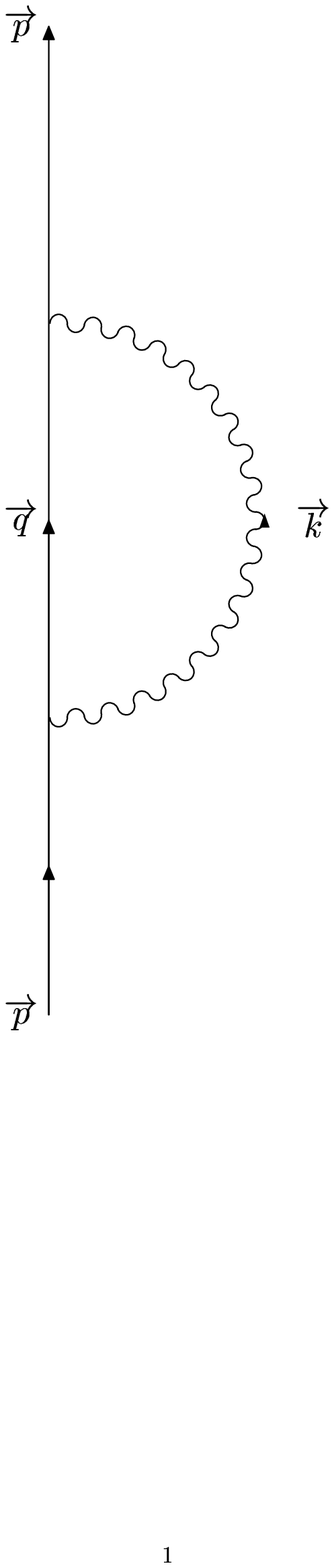}}    
    \caption{The only one loop diagram calculated for radiation
      correction in the noncovariant formalism of QED. At the two
      vertices, either $H^{(1)}$ or $H^{(2)}$ is used (no interference
      between them would occur due to no polarization in the plane
      wave p state). }
    \label{fig:feymann.1}
  \end{center}
\end{figure}

\end{document}